# Analysis of Rotational Motion based on Rolling Friction Torque

Shosuke Sasaki[1], Yohe Namba[2], Tadao Iwanari[3] and Yasuyuki Kitano[4]

[1] Center for Advanced High Magnetic Field Science, Graduate School of Science, Osaka University, Osaka, Japan

[2] Izumo Shimane Prefectural High School, Shimane, Japan

[3] Matsue Minami High School Supplementary Course, Shimane, Japan

[4] Professor Emeritus, Shimane University, Shimane, Japan

E-mail: sasaki@mag.ahmf.sci.osaka-u.ac.jp, zazensou@gmail.com

**Abstract**

In the problem of cylinder rolling without slipping on a horizontal floor, both the cylinder and floor are generally treated as rigid bodies in normal textbooks. When the air resistance is ignored, the equation of motion has a solution with a constant velocity. However, in the real world, permanent motion does not occur. The difficulty cannot be solved only by the horizontal force, because a horizontal force opposite the translational direction increases the angular velocity of rotation around the center.    Therefore other mechanisms need to be examined. There are two main reasons for this result. 1) Both a cylinder and a floor are not perfect circle and perfect plane, but have uneven surfaces. The micro bumps on the surface yield small collisions in the direction perpendicular to the floor. The collisions generate a rolling friction torque around the center. 2) A strong force acts on the contact part which is deformed. The high-speed deformation produces a history effect on the relationship between stress and strain, because the compressed wave in the contact part diffuses to the outside at the speed of sound. Therefore a rolling friction torque is also generated. Both torques are caused by forces perpendicular to the floor. The rolling friction torque eliminates the discrepancy between the textbook results and reality by solving the simultaneous differential equations of rotation and translation. This method is useful for studying rolling systems such as trains and cars.

## 1. Introduction

The motion of a rotating body can be easily observed in many machines. Therefore it is important to establish a basis for describing the rolling motion. In normal physics textbooks [1], [2], [3], both a cylinder and a floor are treated as rigid bodies in the problem of cylinder rolling without slipping on a horizontal floor. The simultaneous differential equations of translation and rotation are solved. Ignoring the air resistance, the solution has a constant velocity and keeps rolling forever. Therefore this handling is inadequate.

To avoid this contradiction with reality, the rolling friction force was introduced into the equation of translational motion while ignoring the rotational motion. However, excluding the rotation equation makes it impossible to analyze the differences in the rotating parts (axle bearing, drive gear, etc.) of each wheel in detail. Various types of torque are applied to the wheel during braking, acceleration, and curve operation in vehicles. Accordingly, ignoring the rotation equation makes detailed analyses difficult in vehicle (mobile) research. This difficulty can be solved by introducing the rolling friction torque (not the force) proposed in this study.

There is an opinion that "such a thing has been known for a long time, and it is common sense that it is completely established as rolling friction force." However, we dare to examine the details. Let us consider a rolling body that travels on a horizontal floor.

In normal handling, the rolling friction force is phenomenologically applied to the contact area between the rolling body and floor. The force is horizontal and opposite to the direction of travel. Looking only at the translational movement, the center of gravity slows down, so there seems to be no problem. However, the torque caused by the rolling friction force accelerates the rotational angular velocity. That is, any force parallel to the floor has the opposite effect (acceleration/deceleration) on translational and rotational movements. This is the root of the permanent motion derived by solving the rolling motion in physics textbooks.

Recently, there has been an increase in research such as dividing a rolling body and a floor into meshes to solve partial differential equations [4]. Therefore the rolling movement is dynamically simulated. We need an easier way to study the movement of the rolling body without performing these complicated calculations.

To resolve the contradiction between permanent movement and reality, it is necessary to consider the vertical force on the floor. The contact stress at rest was investigated by Hertz [5] in 1881. Since then, much research has been conducted in this area [6]. In addition, a different approach was proposed by Yakushiji et al. [7], who considered the collision between a regular polygonal rolling body and the floor. However, these studies lack an analysis of velocity dependence, energy loss, etc.

The vertical force against the floor mainly produces two types of rolling friction torques around the center. One is due to the minute unevenness of the rolling body and floor surface, as discussed in Section 3. The second is due to the historical effect of the stress-strain relationship in the contact area, as shown in Section 4. In Section 5, we solve the simultaneous differential equations of the rotational and translational motions based on the rolling friction torque. This result reveals the basic structure of rolling friction. Because almost all machines have a rotating body built in, the method described here is useful for analyzing rolling motion.

## 2. Handling of rolling motion in normal text books

In many textbooks [1], [2], [3], the rolling motion of a cylinder can be explained by solving simultaneous differential equations of translation and rotation. This section describes the normal handling of textbooks. The conceptual diagram is shown in Fig. 1. The cylinders and floors are treated as rigid bodies.

The origin is the grounding point of the cylinder at time $t = 0$, and the $z$-axis is vertically upward, the $x$-axis is in the travelling direction, and the $y$-axis is oriented from the front to the back of the paper. Let $a$ be the radius of the cylinder, and $\theta$ be the rotation angle from the initial position as shown in Fig. 1. Without slipping motion the relationship between velocity $v$ and angular velocity $\omega$ is given by the following equation.

$$x = a\theta, \qquad v = a\omega \qquad (1)$$

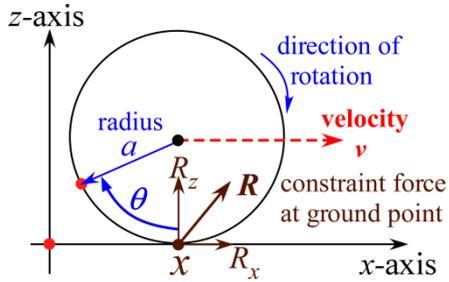

Fig.1 Cylindrical rolling motion without sliding on a horizontal plane

The constraint force $\boldsymbol{R} = (R_x, R_y, R_z)$ acts on the contact point to the floor. Let $m$ be the mass of the cylinder and $g$ be the gravitational acceleration. The relation $R_z = mg$ is derived from the equation of motion in the $z$-direction. The equations for translational movement and rotational movement are expressed as follows,

$$m \frac{d^2 x}{dt^2} = R_x \qquad (2)$$

$$I \frac{d^2 \theta}{dt^2} = -R_x a \quad \text{where } I = \frac{ma^2}{2} \qquad (3)$$

Elimination of $R_x$ from Eqs.(2) and (3) derives–

$$\frac{3m}{2} \frac{d^2 x}{dt^2} = 0 \qquad (4)$$

The solution of Eq.(4) is a horizontal movement at a constant velocity. Substitution of the constant velocity into Eq.(2) yields $R_x = 0$, that is nothing of the $x$ component of $\boldsymbol{R}$. Thus, a permanent motion occurs. Originally, $R_x$ was assumed to exist, but it disappeared. This gives a strange impression. The discomfort is caused by the fact that the rolling friction torque acting on the cylinder from the floor was not considered.

## 3. Rolling friction torque due to unevenness of surfaces in cylinder and floor

In this section, we examine the rolling friction caused by the unevenness on the surface of the cylinder and floor. Until now, there are many investigations for irregularities. A regular polygon with many sides is a typical example [2], [7].

The potential energy of gravity is lowest when the polygonal side is in contact with a horizontal plane. The rolling movement of a polygon requires energy to lift the center of the polygon exactly above the corner. Reference [7] assumed that all lifting energy $W$ is converted to heat in the collision with the floor. Their assumption is inadequate for two reasons: 1) In the collision the vertical velocity of the center of gravity is smaller than the horizontal velocity. Therefore the kinetic energy in the horizontal direction is not lost by the collision. 2) Part of the vertical energy is replaced by heat and the rest becomes recoil kinetic energy. A detailed analysis will be performed as follows.

Figure 2 shows a cross-sectional view of a polygon.

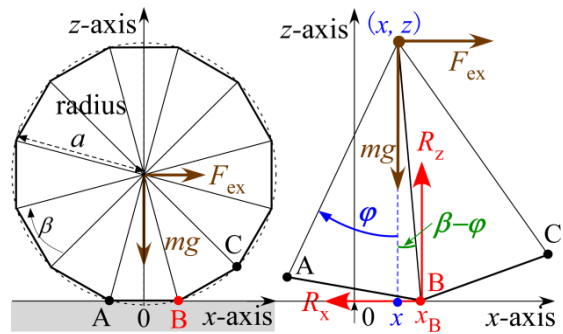

Fig.2 Torque of rolling friction for polygon

Here, as an example, we use a dodecagon with radius $a$ and side central angle $\beta = \pi/6$. All coordinate axes are the same as those in Fig. 1. The right panel in Fig. 2 shows an enlarged figure where the external force $F_{ex}$ acts horizontally at the center and the dodecagon starts to rotate slowly. The rotation angle and the torque are measured clockwise in this study. In the left panel of Fig. 2, the length of the OB is $a \sin(\beta/2)$.

Let $m$ be the mass of the cylinder and $g$ be the gravitational acceleration. The gravitational torque around point B is $-mga \sin(\beta/2)$ at the initial state. The rightward external force $F_{ex}$ produces torque $N_{ex} \approx a F_{ex}$ around B; thus, the maximum torque $N_{exMax}$ without rolling is given by the following equation:

$$N_{exMax} = mga \sin\frac{\beta}{2} \quad (5)$$

When a torque larger than this value is applied, the polygon starts to rotate. The right panel of Fig.2 shows an enlarged view after the rotation starts. Here, the angle $\varphi$ is the sum of the angle $(\beta/2)$ and the rotation angle from the initial state. The length $x_B - x$ is $a \sin(\beta - \varphi)$, and the gravitational torque around point B is $-mga \sin(\beta - \varphi)$. The torque $N_{ex}$ around B due to the external force balances with the gravitational torque under quasi-static motion as

$$N_{ex} = mga \sin(\beta - \varphi) \quad (6)$$

Let $\mathbf{R} = (R_x, R_y, R_z)$ be the force acting on the cylinder from the floor at B. In a quasi-static rotation, the $x$ component $R_x$ balances with the external force, and the $z$ component $R_z$ balances with gravity as follows:

$$R_x = -F_{ex} \quad (7a)$$
$$R_z = mg \quad (7b)$$

Polygons at the three positions are shown in Fig. 3. When moving the center of gravity to just above point B from the initial state, the work $W_1$ done by the external torque $N_{ex}$ is

$$W_1 = \int_{\beta/2}^{\beta} N_{ex} d\varphi = mga(1 - \cos(\beta/2)) \quad (8)$$

This energy is equal to the increase value $\Delta U$ in the potential energy $U$.

$$\Delta U = mga(1 - \cos(\beta/2)) \quad (9)$$

After the center of gravity is exactly above point B, the rotation accelerates. Therefore the vertical and horizontal velocities of gravity center are generated. Thereafter, the polygon collides with floor BC. During the collision, the polygon receives an upward force from the floor.

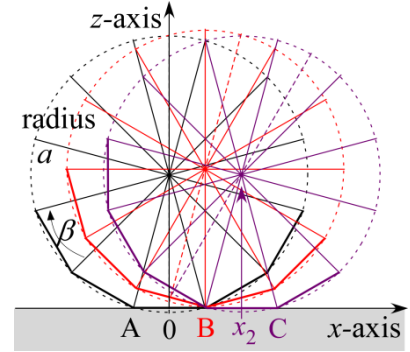

Fig.3 Polygonal rolling (for three positions)

When this system is solved by a perfect elastic collision, the velocity to the right gradually decreases after the collision, and it stops when the center of gravity is directly above point C. Thus, perfect elastic collision will not lose mechanical energy. By contrast, Ref. [7] considers that the polygon loses all kinetic energy and is still at the moment of collision. Both conclusions (no loss, total loss) are out of reality. In fact the vertical velocity of the center of gravity is small compared with the horizontal velocity during a collision. The ratio of vertical velocity to horizontal velocity of gravity center is $\tan(\beta/2)$. Moreover, the contact area between the polygon and the floor is deformed, and only a small part of the kinetic energy is converted into thermal energy. The remaining kinetic energy continues to move the polygon. Because the heat loss varies depending on the materials of the polygon and floor, it is necessary to go into the analysis via materials science.

Our main purpose is to study the rotational motion; accordingly, the ratio of the heat loss part is treated as a parameter. Let $c$ be the ratio of the energy lost during a collision. Then, $c\,W_1$ is the thermal energy caused by the collision. In other words, the mechanical energy of $c\,W_1$ is lost during the rotation angle $\beta$ from the initial position. When $\beta$ is small, the required average torque $N_{\text{exMean}}$ is

$$N_{\text{exMean}} = \frac{c\,W_1}{\beta} \approx mga\,c\,\frac{\beta}{8} \qquad (10)$$

The above result holds even if it is not a regular polygon. The same result can be obtained with only the unevenness of the floor. Let the value of $\beta$ be the central angle formed by the nearest two corners (micro bumps) of the various irregularities existing on the cylinder/floor, the same result will be obtained. Therefore, the coefficients $c$ and $\beta$ should be considered as average quantities determined by the properties of the rotating body and floor. Subsequently, the torque in Eq. (10) is required to roll the cylinder on the floor. Consequently, the rolling friction torque $N_{\text{unevenness}}$ is given by the following equation when $\beta$ is small:

$$N_{\text{unevenness}} \approx -mga\,c\,\frac{\beta}{8} \qquad (11)$$

Let us compare this theoretical value with the experimental value. According to the Mechanical Engineering Handbook [8], the coefficient of sliding friction between iron and iron is 0.52. For the rolling movement, the coefficient of rolling friction of 1/16 in $\phi$ steel ball is $0.000014 \sim 0.001$ when rolling on planes such as glass, iron, lead, copper, or aluminum. Even for a top-quality cylinder or sphere with a radius of approximately 1 mm, the radius deviation from a perfect circle or sphere is more than 1 nm. It is impossible to make a deviation of less than 0.1 nm considering the size of the atom. Because the deviation is equal to $a(1 - \cos(\beta/2))$, the central angle $\beta$ is approximately $2.8 \times 10^{-3}$ rad for radius $a \approx 1$ mm. The coefficient of rolling friction of the steel ball on glass (which seems to have the best flatness) is 0.000014 in the data of the Mechanical Engineering Handbook. The theoretical value of Yakushiji et al. is $\beta/8$ which is the value in the case of $c = 1$ in Eq.(11) (see Ref. [7]).

The theoretical coefficient $\beta/8$ is approximately $3.5 \times 10^{-4}$, and is approximately 25 times the experimental value 0.000014. To explain the experimental value only by the deviation from the perfect circle, the deviation must be less than 0.01 nm. This is impossible because it is smaller than the size of an atom. Therefore, the results of Ref. [7] are necessary to improve as Eq.(11). For small $\beta$, the vertical velocity of the gravity center at the collision is a second-order minute amount compared to the horizontal velocity of the gravity center, even for quasi-static movement. Therefore, the energy loss in an uneven collision (bump collision) is a small part of the kinetic energy.

The rolling friction torque $N_{\text{unevenness}}$ via unevenness is divided by $mga$ and thus the rolling friction coefficient is derived to be $c\,\beta/8$. Assuming that the energy loss ratio in collision is $c \approx 0.01 \sim 1$, $c\,\beta/8$ is about $10^{-3} \sim 10^{-5}$. The order of magnitude matched the experimental values. In an actual steel ball, the deviation seems to be much larger than 1 nm; thus, the rolling friction coefficient depends on the quality of the steel ball.

In the case of automobiles, the road surface has irregularities of approximately 10 mm order, so $\beta$ is as large as 0.5. Therefore, the invention of tires exerts an epoch-making effect [9]. Although there is heat loss caused by rubber deformation due to the unevenness of the road surface, the air inside the tire changes to be almost adiabatic (almost no energy loss). Therefore the energy loss was small. This can be easily understood by comparing it with a deflated tire. Compared to iron and wooden wheels, tires have an extremely small rolling friction coefficient because of the small value of $c$.

In high-speed rolling, the deformation speed is very high at the contact part. Thereby, the deformation energy per second increases and propagates outside at the speed of sound. This effect is analyzed in the next section.

## 4. History effect of relationship in stress and strain

Even if the surface of the cylinder or floor is perfect, the kinetic energy is lost owing to the contact deformation. Before analyzing the process of kinetic energy loss, we searched for reliable measurements of the rolling motion. One of them is the measurement on the Shinkansen, which enables highly reproducible and reliable measurements over a very long distance. The radius of the Shinkansen wheels is 0.43 m. When traveling at 200 km/h, the time that a specific part of the wheel touches the rail is approximately $10^{-4}$ sec, and the time that it does not touch is approximately 0.05 sec. Because the velocity of a longitudinal wave inside the iron is approximately 6000 m/s, the distance traveled in approximately $0.5 \times 10^{-4}$ sec (half the contact time) is approximately 0.3 m. The compression wave of the contact did not reach the center of the axle. The strain/stress history curve for such a high-speed deformation has not been measured. Therefore, we examine the history curve of stress and strain generated by such a short-time deformation by a thought experiment. It can be called desk theory, but the following analysis shows the characteristics of this phenomenon.

The strain state when the rod-shaped steel is compressed at a high speed is shown in the left panel of Fig. 4. The magnitude of the compressive stress is color-coded so that it can be conceptually understood. The real stress distribution is continuous, so it cannot be color-coded, but it is an easy-to-understand diagram. The white part has zero stress, and the stress increases as the black becomes darker. Fix the upper end assuming the center of the wheel. The lower end is the part of the contact with the rail. After contact, the stresses $f_1, f_2, f_3$ with the relation $f_2 = 2f_1$, $f_3 = 3f_1$ is applied to the contact part at times $t_1, t_2, t_3$, respectively. The stress is zero at time $t_0$, the color of the bottom of the rod is white. As shown in Fig.4, stresses $f_1, f_2, f_3, f_2, f_1, 0$ act from the bottom at times $t_1, t_2, t_3, t_4, t_5, t_6$, respectively.

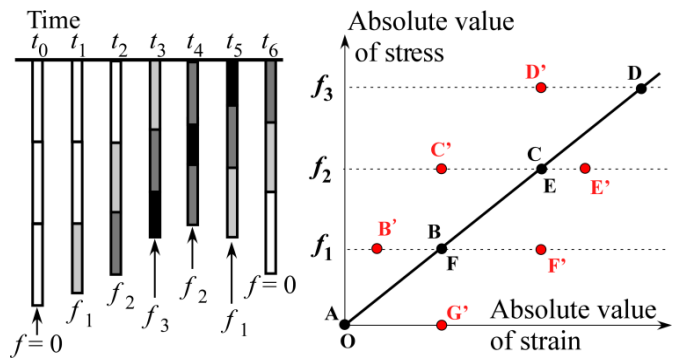

Fig.4  Left panel shows that the top of the rod is fixed and the length of the rod changes due to the stress applied from the bottom. The stress increases from zero to $f_1, f_2 = 2f_1, f_3 = 3f_1$, and then decreases to $f_2, f_1$ and zero. In high speed deformation the compressed part is transmitted upward at the sound speed. The rod color is divided into three parts. Each color represents the stress value. The right panel is a graph of relationship between strain and stress. Points A, B', C', D', E', F', G' correspond to times $t_0, t_1, t_2, t_3, t_4, t_5, t_6$, respectively.

The average strain of the entire rod can be determined by measuring the position of the lower end. Thereby, the strain-stress relationship is shown on the right panel of Fig. 4. The strain-stress relationship at time $t_0, t_1, t_2, t_3, t_4, t_5, t_6$ is represented by points A, B', C', D', E', F', G', respectively. Accordingly the stress increase path does not coincide with the stress decrease path during high-speed deformation. For quasi-static rotation, the stress is uniform throughout the rod. Therefore, the strain-stress relationship is represented by points A, B, C, D, E, F, and A. That is, in the quasi-static deformation, the strain-stress curve exhibits normal elastic deformation. The above analysis assumes that the compressive elastic wave is not reflected from the upper end. This assumption is correct because the actual compression wave in the contact spreads over a wide area of the wheel or rail and does not return to the contact.

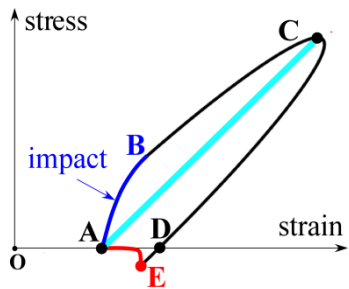

Fig.5　Stress-strain relation in a high speed deformation after long term steady rotation (Residual strain appears at point A ), The contact part between wheel and rail is separated from at D. The straight line AC is the relationship when deformed quasi-statically.

Because the local strain-stress relationship is an elastic process, the contact deformation does not generate heat locally. Of course, the compressed wave spreads and eventually turns into heat. The high-speed deformation discussed here has certain characteristics that differ from the normal history curve in the strain-stress relationship as follows: The gradient value (stress/strain) depends on the deformation speed, for example, lines AB' and AB in Fig.4. The difference (gradient at velocity $v$)-(gradient at zero velocity) is proportional to $v$. This characteristic is consistent with the Shinkansen data examined in Section 7. A detailed analysis will be presented elsewhere. It is a desk discussion, but it is correct for qualitative behavior.

Return to the actual Shinkansen wheel movement. In the initial run, the wheels of the Shinkansen become "familiar state" (material engineering terminology). Then the strain of wheel is remained to be the magnitude of "familiar state." Figure 5 shows the history curve between the strain and stress at high-speed deformation [10]-[13]. The strain at point A remains to be the value of "familiar state" at zero stress. When a particular point on the wheel touches the rail, its strain-stress state begins at point A, as shown in Fig. 5. The contact part receives an impact force, and the strain-stress state rises sharply to point B. After that, the strain and stress increase until approximately $0.5 \times 10^{-4}$ sec, and reach point C. After passing the maximum stress, the strain-stress state reaches point D in approximately $0.5 \times 10^{-4}$ seconds and the specific point on the wheel leaves the rail. Another study showed that the specific point remains in contact with the rail owing to the adhesive force and peels off at point E in Fig.5. In both cases, the strain is relieved within approximately 0.05 seconds (one rotation). At that time, the strain-stress state returns to point A in Fig.5.

Now, let us return to the rotational motion of the cylinder. An enlarged view of the contact area between the cylinder and the floor is shown in Fig.6. The coordinate axes are the same as those shown in Fig. 1. The left panel is stationary and the right panel rotates. Each distribution graph of the normal stress was superimposed on the bottom. The stress from the floor is indicated by upward arrows.

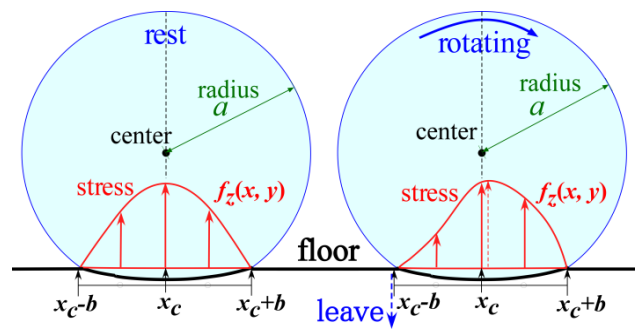

Fig.6　Stress distributions for rest state (left graph) and rolling state (right graph). In the rolling state, the maximum stress position is slightly shifted to the right of $x_c$.

At rest, the stress distribution is symmetrical on the left and right sides of the center of gravity. Therefore, the torque around the center of gravity is zero. When rotation occurs, the stress in the front half of the direction of travel is greater than that at rest owing to the history curve in Fig. 5. On the other hand, in the rear part, it is smaller than at rest. Thus the stress in the front half is larger than that in the rear half. Therefore, the torque around the center of gravity is generated owing to the rolling movement. The rolling friction torque via stress depends on the speed of the deformation, namely the velocity of the rotation. That is, the torque approaches zero as the speed decreases. Such rolling friction torque via stress occurs in any rolling body.

Let $f_z$ be the stress in the $z$ direction. The total amount of stress is obtained by the surface integral, and the integrated value of the stress $f_z$ is balanced with gravity as follows:

$$\iint f_z(x,y)\mathrm{d}x\mathrm{d}y = mg \qquad (12)$$

The function form of the stress $f_z(x,y)$ is shown by the red curve at the bottom of Fig. 6. The stress distribution during rotation is asymmetrical as shown in the right panel. The asymmetric distribution produces a counterclockwise torque $N_{\text{stress}}$ around the center of gravity. The torque $N_{\text{stress}}$ is named "rolling friction torque via stress" which is given by

$$N_{\text{stress}} = -\iint (x - x_\text{C}) f_z(x,y) \mathrm{d}x \mathrm{d}y \qquad (13)$$

As can be seen from Eq. (12), the integration of stress $f_z(x,y)$ is proportional to gravity $mg$. Accordingly $N_{\text{stress}}$ is expressed by employing the dimensionless number $k_{\text{stress}}$, as follows:

$$N_{\text{stress}} = -k_{\text{stress}} mga \qquad (14)$$

In this study the clockwise torque is positive; thus, the coefficient $k_{\text{stress}}$ is a positive value. The history curve of strain stress is shown in Fig. 5. The difference between the increasing stress path and decreasing stress path depends on the rate of deformation. The difference is zero in the quasi-static rolling. Therefore, the coefficient $k_{\text{stress}}$ is composed of terms of the first or higher order of velocity. The coefficient of the rolling friction torque via contact stress is

$$k_{\text{stress}} = k_1 v + k_2 v^2 + \cdots \qquad (15)$$

Shinkansen wheels are re-shaved (mounted on the vehicle body) at a mileage of approximately 100,000 to 600,000 km. Wheel replacement takes place over a mileage of approximately three million kilometer [14]. Therefore, the wheel wear per rotation is extremely low. Although a velocity-independent term appears in the coefficient of the rolling friction torque due to wheel wear, the effect is negligibly small in the Sinkansen data. The velocity-independent term is mainly caused by the surface unevenness (micro bumps) of the wheel and rail as examined in the previous section.

In the result, the total rolling friction torque $N_{\text{friction}}$ is given by the following sum.

$$N_{\text{friction}} = N_{\text{unevenness}} + N_{\text{stress}} + \text{other small effect} \qquad (16)$$

The rolling friction torque coefficient is defined as follows: (coefficient of rolling friction torque):

$$k = |N_{\text{friction}}/(mga)| \qquad (17)$$

Then the total rolling friction torque $N_{\text{friction}}$ depends on the velocity $v$ as shown in Eqs.(11), (14), (15) and (16).

$$N_{\text{friction}} = -mgak = -mga(k_0 + k_1 v + k_2 v^2 + \cdots ) \qquad (18a)$$

where

$$k_0 = (c\beta/8) + (\text{term via wheel wear and so on}) \qquad (18b)$$

The velocity dependence of $k$ was measured in Shinkansen [15], [16]. According to it, the resistance force proportional to gravity is a linear function with respect to the velocity and that of the second order or above can be ignored. This measurement was correct up to approximately 300km/h. Therefore, the "rolling friction torque coefficient" $k$ depends on the velocity in the case of Shinkansen as follows:

$$k \approx k_0 + k_1 v \quad (19)$$

Next, we examine the simultaneous differential equations of rotation and translation in cylinder rolling.

## 5. Equations in rolling motion of cylinder and its solutions

The dynamics of cylinder rolling on a horizontal floor are examined using the rolling friction torque. A schematic illustration of this process is presented in Fig.7. Let the $z$-axis be vertically upward and the $x$-axis be in the direction of horizontal travelling. The $y$-axis is from the front of the page to the back. The origin is the grounding point at the time of departure. The rolling motion is investigated in the case without slippage between the cylinder and the floor.

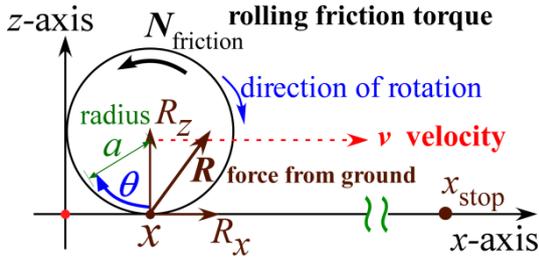

Fig.7 Rolling motion of a cylinder on a floor

Without slippage, the length $a\theta$ is the distance $x$ in Fig.7. Let $t$ be the time and the relationship between the velocity $v$ and the angular velocity $\omega$ is as follows;

$$x = a\theta, v = dx/dt, \omega = d\theta/dt, v = a\omega \quad (20)$$

The constraint force $\boldsymbol{R} = (R_x, R_y, R_z)$ acts on the grounding point of the cylinder. The vertical movement is approximately $a \times (1 - \cos(\beta/2))$ where $\beta$ is the average value of center angles between the nearest micro bumps as mentioned in Section 3. Therefore the vertical movement is negligibly small and the forces in the $z$ direction are in equilibrium as,

$$R_z = mg \quad (21)$$

where $m$ is the mass of the cylinder and $g$ is the gravitational acceleration. In the rolling motion of a cylinder, two differential equations, translational motion and rotational motion, should be satisfied as follows:

(Equation of translational motion): $m\dfrac{d^2 x}{dt^2} = R_x \quad (22)$

(Equation of rotational motion): $I\dfrac{d^2\theta}{dt^2} = N_{\text{friction}} - R_x a$

$$\text{where} \quad I = \frac{ma^2}{2} \quad (23)$$

$I$ is the moment of inertia around the central axis. The elimination of $R_x$ from Eqs.(22) and (23) yields the following equation using Eqs.(20).

$$m\frac{d^2 x}{dt^2} = \frac{ma}{ma^2 + I} N_{\text{friction}} \quad (24)$$

Substitution of Eq.(19) into (18a) yields the rolling friction torque $N_{\text{friction}}$ as follows:

$$N_{\text{friction}} = -mga(k_0 + k_1 v) \quad (25)$$

The coefficients $k_0, k_1$ are determined from the properties of the cylinder and the floor. Eqs.(24) and (25) yield;

$$m\frac{dv}{dt} = -\frac{2mg}{3}(k_0 + k_1 v) \quad (26)$$

Since Eq.(26) is a first-order differential equation with separable form, the solution is obtained as

$$\frac{1}{k_1}\log[(k_0 + k_1 v)] =$$
$$-\frac{2g}{3}t + (\text{integration constant}) \quad (27)$$

Let $v_0$ be the initial velocity at $t = 0$, and the integration constant is obtained. Therefore, the function form of $v$ is

$$(k_0 + k_1 v) = (k_0 + k_1 v_0)\exp\left(-\frac{2gk_1}{3}t\right) \quad (28)$$

The stopping time $t_{\text{stop}}$ is obtained by substituting $v = 0$ into Eq.(28).

$$t_{\text{stop}} = \frac{3}{2gk_1}\log\left(1 + \frac{k_1}{k_0}v_0\right) \quad (29)$$

Integrating both sides of Eq.(28) and using the initial condition $x = 0$ at $t = 0$ gives

$$x = -\frac{k_0}{k_1}t - \left(\frac{k_0}{k_1} + v_0\right)\frac{3}{2gk_1}\left[\exp\left(-\frac{2gk_1}{3}t\right) - 1\right] \quad (30)$$

The distance $\ell$ until the stop can be obtained by substituting $t = t_{\text{stop}}$ into Eq.(30) as,

$$\ell = -\frac{k_0}{k_1}\frac{3}{2gk_1}\log\left(1+\frac{k_1}{k_0}v_0\right) + \frac{3}{2gk_1}v_0 \quad (31)$$

The constraint force $R_x$ is derived by removing the acceleration from Eqs.(22) and (26) as follows:

$$R_x = -\frac{2mg}{3}(k_0 + k_1 v) \quad (32)$$

The force $R_x$ is negative and the torque $N_{\text{friction}} - R_x a$ is also negative. Therefore the angular velocity decelerates from Eq.(23). Consequently the rolling friction torque is required to slow down the rotation.

It is necessary to confirm the condition of no slipping between the cylinder and floor. The constraint force $R_x$ acts between the cylinder and floor; therefore, the value $|R_x|$ should be less than the maximum sliding friction force to avoid slippage. The following condition was required.

$$\frac{2mg}{3}(k_0 + k_1 v) <$$
$$mg \times (\text{maximum sliding friction coefficient}) \quad (33)$$

The left-hand side of inequality (33) is the maximum at the initial velocity; therefore, the next condition is required.

$$\frac{2}{3}(k_0 + k_1 v_0) <$$
$$(\text{maximum sliding friction coefficient}) \quad (34)$$

The initial velocity cannot exceed a certain value to prevent slippage

In conclusion, the contradiction between ordinary textbooks and reality is resolved by the rolling friction torque caused by the force perpendicular to the floor.

## 6. Method to measure the coefficient of rolling friction torque

Figure 8 shows that an inclined plane is connected to the horizontal plane through a region with curvature greater than the radius of the cylinder. On such a slope, the cylinder can roll smoothly from the slop onto the horizontal plane. It is examined the case in this section where the term $k_1 v$ is negligibly small in the coefficient of rolling friction torque. The $x$-axis is the horizontal right direction and the $z$-axis is the vertically upward direction. Let $(x,z)$ be the position of the center of gravity of the cylinder. The $z'$-axis is perpendicular to the inclined plane and the $x'$-axis is paralell to the slope as shown in Fig.8.

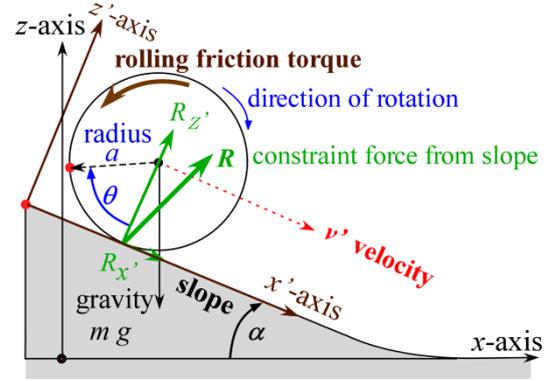

Fig.8 Method to measure the coefficient of rolling friction torque

Let $(x', z')$ be the position of the center of gravity measured in the coordinates of $x'$ and $z'$. We examine the motion in an area with a constant inclined angle $\alpha$. Ignoring the $z'$ direction movement due to the unevenness of the surface, the equation of motion for $z'$ direction is given by

$$m\frac{d^2 z'}{dt^2} = 0 = R_{z'} - mg\cos\alpha \quad (35a)$$

Therefore the normal force $R_{z'}$ is equal to $mg\cos\alpha$. The equation of the translating motion in the $x'$ direction is given by

$$m\frac{d^2 x'}{dt^2} = R_{x'} + mg\sin\alpha \quad (35b)$$

Because the rolling friction torque is the product of the normal force $R_{z'}$ and $ka$, the equation for the rotating motion is

$$I\frac{d^2\theta}{dt^2} = -R_{x'}a - k_0 mga\cos\alpha \quad (36)$$

where $k \approx k_0$ is used. The elimination of $R_{x'}$ from Eqs (35b) and (36) yields the following equation;

$$\left(m + \frac{I}{a^2}\right)\frac{d^2 x'}{dt^2} = mg\sin\alpha - k_0 mg\cos\alpha \quad (37)$$

Let $h(t)$ be the height of the center of gravity at time $t$. The height was $h_0$ in the initial state. The initial position

is $x'(0) = 0$. Then, the height $h(t)$ is related to $x'(t)$ as follows:
$$h(t) = h_0 - x'(t) \sin \alpha \quad (38)$$

The total kinetic energy is the sum of the translational energy $(1/2)mv'^2$ of the center of gravity and the rotational energy $(1/2)I\omega^2$. The mechanical energy $E$ is the sum of the kinetic energy and potential energy $mgh$. So we get

$$E = \left(\frac{1}{2}mv'^2 + \frac{1}{2}I\omega^2\right) + mgh \quad (39)$$

The condition of non-slippage is expressed by $v' = a\omega$ on the slope; therefore, the time differentiation of Eq.(39) becomes

$$\frac{dE}{dt} = \left(mv'\frac{dv'}{dt} + \frac{I}{a^2}v'\frac{dv'}{dt}\right) + mg\frac{dh}{dt} \quad (40)$$

Substitution of Eq.(37) into Eq.(40) yields
$$\frac{dE}{dt} = v'(mg\sin\alpha - k_0 mg\cos\alpha) + mg\frac{dh}{dt} \quad (41)$$
Time derivative of Eq.(38) is $dh/dt = -v'\sin\alpha$ and thus, Eq.(41) becomes the following simple form:
$$\frac{dE}{dt} = -v' k_0 mg \cos\alpha \quad (42)$$
Substitution of $x = x' \cos\alpha$ yields
$$\frac{dE}{dt} = -\frac{dx}{dt} k_0 mg \quad (43)$$
Equation (43) is integrated and yields the following relation.
$$E_1 - E_2 = -k_0 mg(x_1 - x_2) \quad (44)$$
The mechanical energies $E_1, E_2$ and the coordinates $x_1, x_2$ are the values at times $t_1, t_2$ respectively. Equation (44) indicates that the decrease in mechanical energy is proportional to the horizontal component of the distance travelled. The relationship does not depend on the angle of the slope, so it can be used anywhere.

Let $h_0$, $h_{stop}$ be the heights at time $t_0, t_{stop}$, and $x_0, x_{stop}$ be the horizontal coordinates at $t_0, t_{stop}$. Because the velocity is zero at the initial time $t_0 = 0$ and zero at the stopping time $t_{stop}$, Eq. (44) becomes

$$mgh_0 - mgh_{stop} = -k_0 mg(x_0 - x_{stop}) \quad (45a)$$
Consequently, the rolling friction torque coefficient $k_0$ is determined as

$$k_0 = (h_0 - h_{stop})/(x_{stop} - x_0) \quad (45b)$$
In other words, the rolling friction torque coefficient is obtained by dividing the height difference by the moving horizontal distance. Since Eq.(45b) is derived from an arbitrary moment of inertia $I$; it holds even for a sphere or cylinder. Thus we have presented an example of determining the rolling friction torque coefficient.

## 7. Torque required for cases of acceleration/ inertia/ deceleration

In this section, we examine several types of torque and constraint forces in a vehicle. An example is shown in Fig.9. In trains and automobiles, the rotational force of the motor or engine is transmitted to the wheels. Let $N_{driving}$ be the driving torque for each wheel. This drive torque accelerates the vehicle's body. When operating the brakes, braking torque $N_{braking}$ is applied to each wheel to decelerate the vehicle body. Each train has eight or more wheels, and each car has four or more wheels. The total number of wheels is represented by $J$. Let $j$ ($j = 1,2,\cdots J$) be the number specifying the individual wheel. The driving torque acts on some wheels but does not act on the other wheels. As an example, we examine the following case where the drive torque $N_{driving}$ acts on the wheels of $j = 1,2,\cdots (J/2)$ and does not act on the wheels of $j = (J/2) + 1, \cdots, J$.

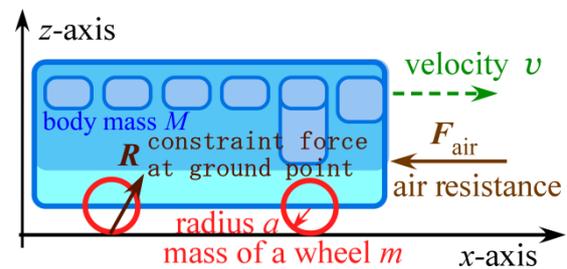

Fig.9 Simple example of vehicle. There are two kinds of wheels, one is driving wheel and the other is inertia running wheel

There are other torques such as rolling friction torque $N_{\text{friction}}$ working on each wheel from the floor (rail), and rotational friction torque $N_{\text{bearing}}$ of the axle bearing. To simplify the problem, we assume that the same magnitude of gravity acts on each wheel. In addition, $N_{\text{friction}}$ and $N_{\text{bearing}}$ are the same for all wheels. Let $M$ be the mass of the vehicle body excluding the wheel mass, and let $m$ be the mass of the wheel. The mass $m$ is the same for all the wheels. Then the total mass is

$$\text{(total mass)} = M + J\,m \quad (46)$$

The horizontal constraint force $R_x^j$ acts on the $j$-th wheel to prevent slippage between the wheel and the floor (rail). Then the equation of motion for the center of gravity is

(equation of motion of the center of gravity):

$$(M + J\,m)\frac{d^2 x}{dt^2} = -F_{\text{air}} + \sum_{j=1}^{J} R_x^j \quad (47)$$

Therein $F_{\text{air}}$ (a positive value) represents the magnitude of air resistance and acts directly on the vehicle body.

*7.1 Equation of rotation*

We examine the rotation motion of each wheel. Let $N_j$ be the total torque acting on the $j$-th wheel around the axle center. The value of $N_j$ depends on operations such as driving, braking and inertia running operations. In the driving case (accelerating drive and constant speed drive), $N_j$ with $1 \le j \le (J/2)$ includes the torque $N_{\text{driving}}$, whereas the others include nothing of $N_{\text{driving}}$.

(driving case) $\quad N_j = N_{\text{friction}} + N_{\text{bearing}} + N_{\text{driving}}$
$\qquad$ (for $1 \le j \le (J/2)$: drive wheel) $\quad$ (48a)

(driving case) $\quad N_j = N_{\text{friction}} + N_{\text{bearing}}$
$\qquad$ (for $(J/2) + 1 \le j \le J$: non-drive wheel) $\quad$ (48b)

(braking case) $\quad N_j = N_{\text{friction}} + N_{\text{bearing}} + N_{\text{braking}}$
$\qquad$ (for all wheels) $\quad$ (48c)

(inertia running case) $\quad N_j = N_{\text{friction}} + N_{\text{bearing}}$
$\qquad$ (for all wheels) $\quad$ (48d)

As shown in Eqs.(48a)-(48d), the same $N_{\text{bearing}}$ value acts on each wheel and the same $N_{\text{braking}}$ value acts on each wheel. The analysis in this section applies the same gravity $(m + M/J)g$ to each wheel; therefore, $N_{\text{friction}}$ takes the same value for all wheels owing to Eq.(25):

$$N_{\text{friction}} = -\left(m + \frac{M}{J}\right) g a (k_0 + k_1 v) \quad (49)$$

In the actual analysis, there are other torques caused by gears transmitting power, by energy recovery type brakes and so on. These torques have specific dependencies on the speed and weight. The torques with these individual characteristics can be added to the right-hand side of Eqs. (48a)-(48d). In this way, the method of this study can incorporate various features according to the actual situation.

The moment of inertia around the center is assumed to have the same value $I$ for all the wheels. Let $\theta_j$ be the rotation angle of the $j$-th wheel. Then the equation of motion around the axle center is

(Equation of rotation around the axle center)

$$I \frac{d^2 \theta_j}{dt^2} = N_j - R_x^j a \quad \text{for } j = 1, 2, \ldots, J \quad (50a)$$

Let us consider straight running which is not a curve. The position $x$ of the center of gravity and the rotation angle $\theta_j$ are taken to be zero at the initial time, and thus a common rotation angle $\theta$ can be introduced as $\theta_j = \theta$. The non-slippage condition yields $x = a\theta$. Accordingly Eq.(50a) becomes

$$\frac{I}{a^2} \frac{d^2 x}{dt^2} = \frac{1}{a} N_j - R_x^j \quad \text{for } j = 1, 2, \ldots, J \quad (50b)$$

The elimination of $\sum_{j=1}^{J} R_x^j$ from equations (47) and (50b) yields the following equation:

$$\left(M + J\,m + J \frac{I}{a^2}\right) \frac{d^2 x}{dt^2} = \frac{1}{a} \sum_{j=1}^{J} N_j - F_{\text{air}} \quad (51)$$

Thus, the equation of motion of the center of gravity was determined. In particular, the right-hand side of Eq. (51) represents the resistance force during the inertial operation as shown in Eq.(48d).

*7.2 Estimation of resistance forces and torques for an example*

As an example, we estimate the air resistance force $F_{\text{air}}$, horizontal constraint force $R_x^j$ and rolling friction torque $N_{\text{friction}}$ by employing the experimental results of the Shinkansen. According to References [15], [16], the force acting on the vehicle has a mechanical resistance part $F_{\text{mechanical}}$ and an air resistance part $F_{\text{air}}$. $F_{\text{mechanical}}$ is proportional to the total mass and $F_{\text{air}}$ is independent of mass. Therefore, $F_{\text{mechanical}}$ is related to the rolling friction torque $N_{\text{friction}}$ as examined in Eq. (18a) in Section 4.

$F_{\text{air}}$ is roughly the sum of the following two terms: one is proportional to the cross-sectional area of the vehicle ($S_1 = $ width $\times$ height), and the other is proportional to the area of the top, bottom, left, and right ($S_2 = $ (width + height) $\times 2 \times$ train length). Furthermore $F_{\text{air}}$ is proportional to the square of the train speed. (There are other factors such as the shape of the leading vehicle and the cover shape of the device under the bottom, but they are omitted here.) The air resistance is given by

$$F_{\text{air}} = (\mu_1 S_1 + \mu_2 S_2) v^2 \qquad (52)$$

where $\mu_1$ and $\mu_2$ are the coefficient values peculiar to the train, and $F_{\text{air}}$ represents the magnitude (positive value). $F_{\text{air}}$ acts directly on the train in the direction opposite to the travelling.

Next, we examine the magnitude of $N_{\text{bearing}}$. According to Ref. [17], axle bearings mainly have double-row tapered roller bearings or double-row cylindrical roller bearings inside an outer diameter of $\phi$ 210 to $\phi$ 250 mm. Axle bearings have been improved specifically for railway vehicles and the frictional resistance is reduced even for a heavy weight of trains. Furthermore, the actual rotating part is a small "roller" with a radius on the order of cm (10mm). Because the torque is proportional to the radius and the radius of the wheel is several tens of times the radius of the roller, the rolling friction torque due to the "roller" is negligibly small compared to the rolling friction torque between the wheel and the rail.

**(Consideration of the resistance force during inertia running)**

In an inertia running, all the wheels receive the same torque as in Eq. (48d), the right side of Eq. (51) is

$$(\text{Resistance force}) = \frac{J}{a}\left(N_{\text{friction}} + N_{\text{bearing}}\right) - F_{\text{air}}$$
$$\text{for inertia running} \quad (53a)$$

As mentioned above, $N_{\text{bearing}}$ is small; therefore, we ignore it in a simple estimation of this section. In addition, both sides of Eq.(53a) are divided by the total gravity and the result yields the relationship of the dimensionless number as

$$\left|\text{Resistance force}/((M + J\ m)g)\right| = (k_0 + k_1 v) + k_{\text{air}} v^2 \qquad (53b)$$

where $\quad k_{\text{air}} = \frac{(\mu_1 S_1 + \mu_2 S_2)}{(M+J\ m)g} \quad (53c)$

The coefficient values $k_0$, $k_1$, $k_{\text{air}}$ are estimated from the Shinkansen data as an example. The train weight of the Shinkansen 700 series is approximately 600 tons as shown in Fig. 4 of Ref. [15]. Figure 3 in Ref. [15] shows the velocity dependence of mechanical resistance and air resistance forces. The coefficients $k_0, k_1, k_{\text{air}}$ are determined from the graph as follows:

$$k_0 \approx \frac{10}{600 \times 9.8} \approx 0.0015 \qquad (54a)$$
$$k_1 \approx 0.000015 \tfrac{\text{hour}}{\text{km}} \qquad (54b)$$
$$k_{\text{air}} \approx 0.00000011 \left(\tfrac{\text{hour}}{\text{km}}\right)^2 \qquad (54c)$$

Next, the $x$ component of the constraint force $R_x^j$ for $j$-th wheel is calculated by introducing a dimensionless number $\eta$ as follows;

$$\eta = \frac{I}{Ma^2 + J\ ma^2 + J\ I} \qquad (55)$$

Multiplying both sides of Eq. (51) by $\eta$ and subtracting both sides of Eq. (50b) yields the following relationship:

$$R_x^j = \frac{1}{a} N_j - \frac{\eta}{a} \sum_{j=1}^{J} N_j + \eta F_{\text{air}} \text{ for } j = 1, 2, \dots, J \quad (56)$$

Equation (56) indicates that $R_x^j$ depends on the type of train operation because Eqs.(48a)-(48d).

## 7.3 Velocity dependence of driving torque and constraint forces

Let us calculate the driving torque and horizontal constraint force $R_x^j$ in the three types of train operations: (a) constant-speed operation, (b) acceleration operation, and (c) deceleration operation.

### a) Constant speed operation:

The left side of Eq. (51) is zero in constant speed operation; therefore, $\sum_{j=1}^{J} N_j/a = F_{\text{air}}$. Because the present example has $J/2$ driving wheels and $J/2$ inertia wheels, $\sum_{j=1}^{J} N_j$ becomes $J\left(N_{\text{friction}} + (1/2)N_{\text{driving}}\right)$ by omitting the small term $N_{\text{bearing}}$. Use of $F_{\text{air}} = (M + J\,m)g k_{\text{air}} v^2$ yields

$$J \frac{(1/2)N_{\text{driving}} + N_{\text{friction}}}{a} = (M + J\,m)g k_{\text{air}} v^2 \quad (57a)$$

By dividing both sides of Eq.(57a) by $(M + J\,m)g/2$ yields a dimensionless quantity. Therefore the re-expression of $N_{\text{friction}}$ using Eq.(49) yields the following result:

$$\frac{N_{\text{driving}}}{a((M/J)+m)g} = 2(k_0 + k_1 v) + 2k_{\text{air}} v^2 \quad (57b)$$

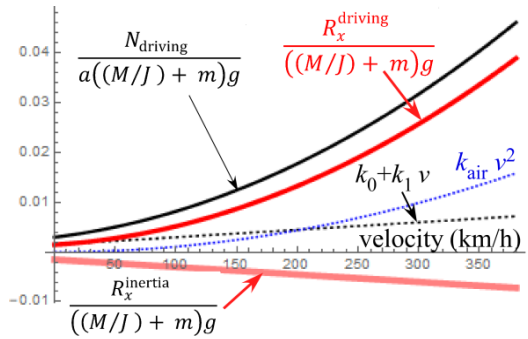

Fig.10 Speed dependence of driving torque and horizontal constraint forces of driving and inertia wheels for constant speed operation

Next, two horizontal constraint forces are calculated. One is represented as $R_x^{\text{driving}}$ which is equivalent to $R_x^j$ acting on the driving wheels for $1 \leq j \leq J/2$. The second is $R_x^{\text{inertia}}$ for $(J/2)+1 \leq j \leq J$.

Substitution of Eq.(57b) into Eq.(50b) and ignoring $N_{\text{bearing}}$ yields the following relations:

$$\frac{R_x^{\text{driving}}}{((M/J)+m)g} = (k_0 + k_1 v) + 2k_{\text{air}} v^2 \quad (58a)$$

$$\frac{R_x^{\text{inertia}}}{((M/J)+m)g} = -(k_0 + k_1 v) \quad (58b)$$

Figure 10 shows three speed dependences of the drive torque and horizontal constraint forces as given in Eqs.(57b), (58a) and (58b). From Fig. 10, the direction of $R_x^{\text{driving}}$ is different from the direction of $R_x^{\text{inertia}}$.

### b) Accelerating operation

The acceleration at the departure of the Shinkansen (Nozomi) is approximately 2.6 km/hour/s, which is approximately 0.074 times the gravitational acceleration. In this case, the substitution of $d^2x/dt^2 \approx 0.074\,g$ into Eq.(51) derives the following equation:

$$\left(M + J\,m + \frac{JI}{a^2}\right)0.074 g = \frac{J}{a}\left(\frac{1}{2}N_{\text{driving}} + N_{\text{friction}}\right) - F_{\text{air}}$$

To simplify the calculation, the moment of inertia is omitted, $\eta = 0$, and $N_{\text{bearing}}$ is also ignored because of its small value. This approximation is named "A-1" hereafter.

Under the approximation A-1, $N_{\text{driving}}$ depends on the velocity as:

$$\frac{N_{\text{driving}}}{((M/J)+m)ga} = 2 \times 0.074 + 2(k_0 + k_1 v) + 2k_{\text{air}} v^2 \quad (59)$$

The horizontal constraint forces $R_x^{\text{driving}}$ and $R_x^{\text{inertia}}$ are derived from Eqs. (56) under approximation A-1 as:

$$\frac{R_x^{\text{driving}}}{((M/J)+m)g} = 2 \times 0.074 + (k_0 + k_1 v) + 2k_{\text{air}} v^2 \quad (60a)$$

$$\frac{R_x^{\text{inertia}}}{((M/J)+m)g} = -(k_0 + k_1 v) \quad (60b)$$

Therefore, the direction of $R_x^j$ is opposite between the driving wheel and the inertia wheel.

### c) Decelerating operation

We examine the case where the acceleration is $(-Z)$ times the acceleration of gravity. Then, the braking torque becomes

$$\frac{N_{\text{braking}}}{((M/J)+m)ga} = -Z + (k_0 + k_1 v) + k_{\text{air}} v^2 \quad (61)$$

The horizontal constraint forces $R_x^j$ are obtained by substituting Eq.(61) into Eq.(56) and then $R_x^j$ is the same for all wheels.

Although this paper has not dealt with curve driving, the following short comment is added. The train wheel has a large inner radius and a small outer radius. On a curve, the outer wheel contacts the rail at a large radius and the inner wheel contacts the rail at a small radius. That is, the wheel radii are different from each other. In automobiles, differential gears work on curves, and the number of rotations of the left and right wheels changes. When examining the movement on a curve, Eqs.(50a) and (50b) should be modified so that the radius or rotation angle depends on the identification number $j$ of each wheel. In this way a detailed analysis can be performed for curve driving.

The magnitude of $R_x^j$ obtained using Eqs.(56) should be less than the maximum sliding friction force between the wheel and rail to prevent slippage. The restriction varies depending on the train operation and speed.

## 8. Conclusion

We have considered the dynamics of a rotating body rolling on a horizontal floor. The rolling friction torque is generated by several causes, two of which provide the main contribution. The first is caused by the minute unevenness on the surfaces of the rotating body and floor which is $N_{\text{unevenness}}$. The second comes from the history effect in the stress-strain relationship at the contact area because the stress dissipates to the periphery at the speed of sound, and the energy does not return to the contact area. The dissipated energy is greater at higher speeds, and thus the history effect becomes greater at higher speeds. Accordingly the rolling friction torque $N_{\text{stress}}$ via the history effect is proportional to the velocity. The rolling friction torque is mainly the sum of $N_{\text{unevenness}}$ and $N_{\text{stress}}$. Based on this rolling friction torque, the fundamental law of rolling motion is expressed by the simultaneous differential equations of rotation and translation. The coefficient of the rolling friction torque was well defined. In addition, an example of how to measure the value is shown at low speed.

The horizontal constraint force $R_x^j$ in the contact area is derived from the non-slippage condition between the wheel and floor (rail or road surface). The direction of $R_x^j$ depends on the conditions of each wheel. There are cases where the direction of $R_x^j$ is the travelling direction for some wheels but is opposite to that of the others. This point is significantly different from the fact that the sliding frictional force always points in the direction opposite to the sliding direction of the wheel.

In conclusion, by faithfully following the basic law of physics and analyzing the equations of motion based on the rolling friction torque, the rotation of each wheel can be understood in detail. There are many rotating parts in machines; therefore, the method developed in this study provides useful means.

### Acknowledgements

The authors would like to thank Professor K. Fujii and Professor T. Iwamoto for their valuable suggestions.